\documentclass{article} 
\usepackage{nips15submit_e,times}
\usepackage{hyperref}
\usepackage{url}
\usepackage{graphicx}
\usepackage{amsmath}
\usepackage{amssymb}
\usepackage{subfig}

\title{System for Filtering Messages on Social Media Content}
\author{
Jinju Joby P. \\
Department of Computer Science and Engineering \\
Christ University Faculty of Engineering \\
Bangalore, India \\
\And
Jyothi Korra\\
Computer Science and Engineering\\
Christ University Faculty of Engineering\\
Bangalore, India\\
\texttt{jyothi.korra@christuniversity.in} \\
}

\nipsfinalcopy 
\begin{document}
\maketitle

\begin{abstract}
The social networking era has left us with little privacy. The
details of the social network users are published on Social
Networking sites. Vulnerability has reached new heights due
to the overpowering effects of social networking. The sites
like Facebook, Twitter are having a huge set of users who
publish their files, comments, messages in other users walls.
These messages and comments could be of any nature. Even
friends could post a comment that would harm a persons
integrity. Thus there has to be a system which will monitor the
messages and comments that are posted on the walls. If the
messages are found to be neutral (does not have any harmful
content), then it can be published. If the messages are found to
have non-neutral content in them, then these messages would
be blocked by the social network manager. The messages that
are non-neutral would be of sexual, offensive, hatred, pun
intended nature.
Thus the social network manager can classify content as
neutral and non-neutral and notify the user if there seems to
be messages of non-neutral behavior.
\end{abstract}

\section{Motivation}

Social Networking Services provide a platform for the group of users or more commonly called as the social group, to share real life experiences and communicate with each other. Social networks are public profiles which allow individuals to create a profile, to create a list of users with whom they share connections \cite{REF07}. Most of these web-based services involve e- mail or instant messaging. Social networking allows people of similar interests to share ideas, photos, posts, comments, activities with the other people in their groups. Some social network is also used to complete a task like textEagle\cite{dg4}.
Popular social networking sites would include Facebook, twitter, LinkedIn, Instagram, YouTube, Pinterest \cite{REF05} and such. The other lesser known social networking sites from different countries like Russia and Canada would include Nexopia, Badoo, Bebo, Delphi, SkyRock, Hi5, Orkut, etc. All the above mentioned sites would have the capability to engulf a social community of millions of people. The resources and network capabilities would vary based on the user community.
Along with the positive effects of social networking, it is unavoidable to ignore the negative effects. It is the era of cybercrimes and cyber terrorism. Various attacks are diminishing large organizations which were once stable. The role of social networking is a huge part in this epic change. The privacy and openness factors of an individuals life have drastically changed over time. Social network asks user to get permission if someone tags a photo of a user. There are techniques where some social networks masks uses\cite{dg6}. Thus it becomes easier for an attacker to target the vulnerable areas of the organization by getting information about an individual.
This would bring light to how the social networking culture needs to be secured. The purpose of having a secure face in the society plays a key role in avoiding any changes of attacks that would diminish or tarnish an individuals social face. 
\begin{figure}
\centering
\includegraphics{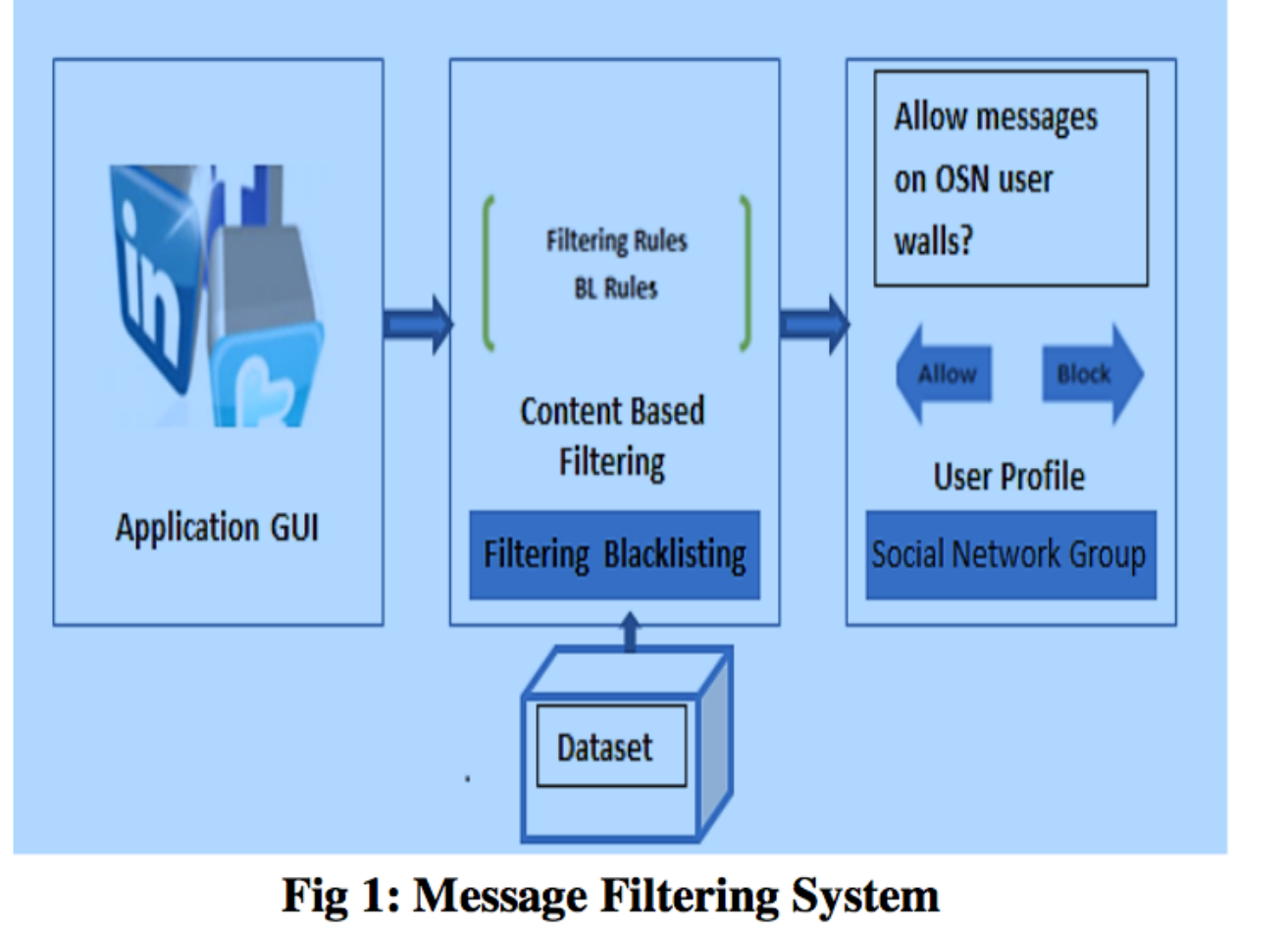}
\end{figure}
\section{Related Work}
\subsection{Content based Filtering}
The architecture in support of OSN services is a three-tier structure \cite{REF01} \cite{REF03}.
Application Graphical User Interface: It is any application that allows users to interact with other users. It can be the social networking sites or electronic mail or instant messaging. Designing the visualization and behavior of a GUI is an important aspect for human-computer interaction. Its goal is to enhance the efficiency and improve the usability of the application for the user. Methods of user-centered design are used to ensure that the visual language used in the application is well accepted by the users. The message filtering system that was described by authors who have worked on text filtering of message content was analyzed to understand the different components. Another important factor to remember is that user scan messages from top-down \cite{dg5}. 
Social Network Manager: The Manager is responsible for handling all the traffic and network and all other operations done by users on the social network. The task of the manager is offline and the social network manager would not be involved in the live message feeds. His responsibilities are to provide the basic OSN functionalities like profile and relationship management. Hence the Social Network Manager can be internal and external.
Social Network Applications: Some OSNs provide an additional layer allowing the support of external Social Network Applications (SNA). The users interact with the system by setting up filtering rules, according to which messages have to be filtered out. Moreover, the GUI provides users with a filtered wall, where only messages that are authorized according to their filtering rules are published.

\section{NAÏVE BAYES CLASSIFICATION}
The Naïve Bayes Classification is done using the Bayes theorem. The Bayes theorem is as follows,
$$P(A/B) = \frac{P(B/A)P(A)}{P(B)}$$
Naive-Bayes classifier trains data to estimate the probability of each item in a document and to further identify and classify a new instance of an item. The equation defined below is the posterior probability that the test theory is true with respect to its presence in the document \cite{REF06}.
$$P(T/T_i) = \frac{P(T_i/T) P(T)}{P(T_i/T) P(T) + P(T_i/\neg T)P(\neg T)}$$
where T is the test theory, Ti is the existing theory in the document. P ( ) defines the probability function. The above formula is to find the probability of presence of a word in a document.
The probabilities of each combination of the sets of columns would be calculated. The classification can be done using the values obtained from the probability function which was mentioned earlier. Thus the classes for each user based on the content in the messages would be calculated. Based on these values the users would be placed in the restricted category as
mentioned. They can fall into the non- neutral classes which can be sexual, hatred, offensive, pun-intended.
The classes would not contain the users classified. To be noted that a user can fall into different classes at a time depending on the type of messages he wants to post.
The next part of the process would be the prediction. The prediction would necessarily begin when a new comment is added and it needs to be classified. The social network manager would classify the comment into the five classes (neutral, sexual, hatred, offensive, pun-intended). This is done using the training set and learning process by the algorithm. Thus the probability that the messages fall under the neutral or non-neutral category would be from the outcome of the probability function.
As set, the threshold for tolerance is 0.3. So the messages of users who have a probability function outcome to be less that 0.3 are trusted by the social network manager. And these messages would be granted permission to be published on the social wall. Whereas the messages or the users who have the outcome of the probability function more than or equal to 0.3 are marked as non-neutral and would be alarmed to the social network manager. Thus these messages or the user would be blocked from publishing any messages on the user wall.
Based on the necessities of the user, an application can be developed in the desired environment and can be made live on social media sites that we already use.

\section{Implementation}
\subsection{Dataset}
The dataset for this particular experimentation has to be of textual nature. As the area of study is social media data like comments and other information that is displayed in public, the dataset will be a collection of comments in English language as seen in figure 2. Various datasets of this type are available online. This particular dataset was available at the link. It was earlier used for similar research purposes.
http://www.marcovanetti.com/pages/wmsnsec/wmsnsec1.xml The dataset was available in Italian language.
\begin{figure}
\centering
\includegraphics[width=5in,height=3.5in]{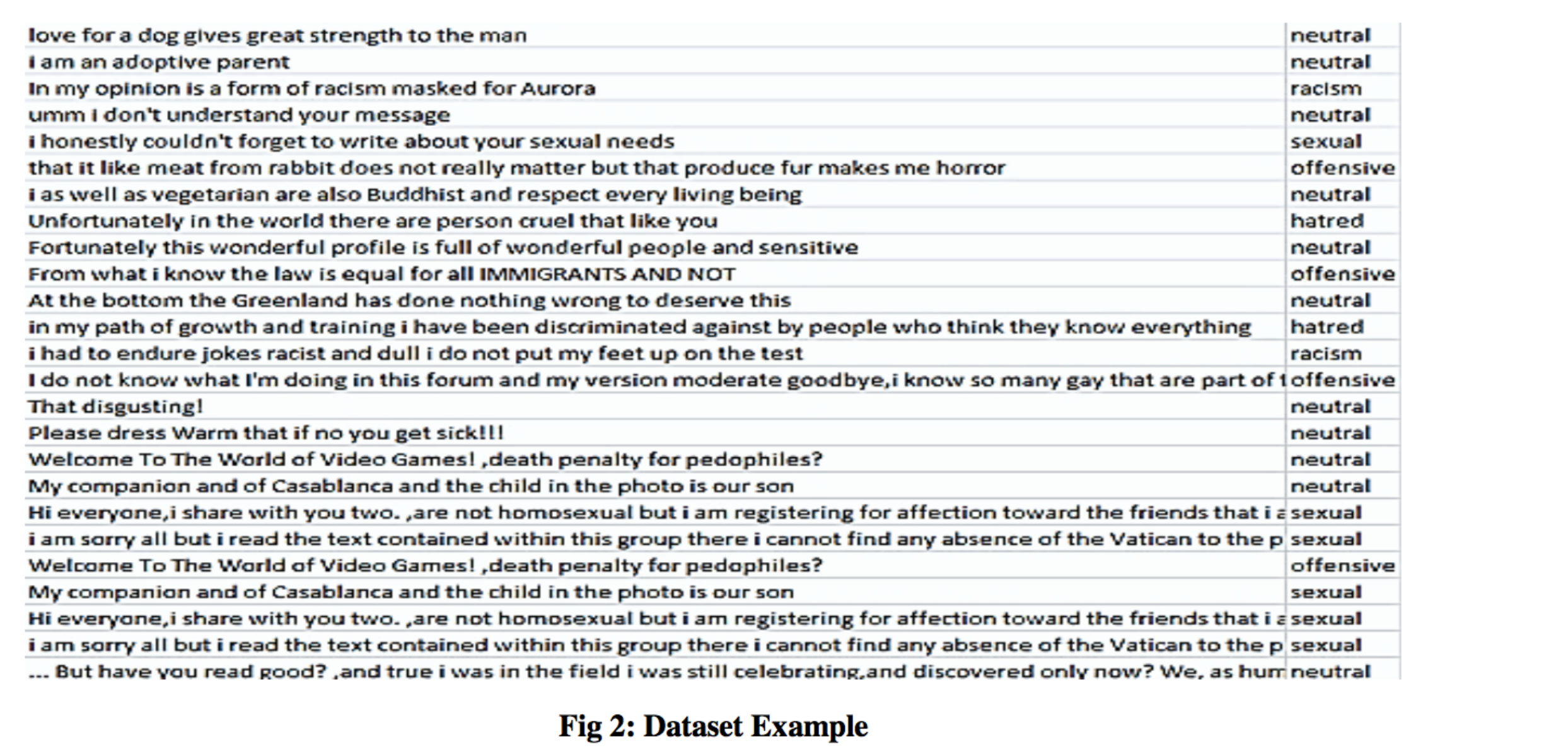}
\end{figure}
\subsection{Preprocessing}
The raw form of the dataset cannot be used as the input to the classifier as it would contain stop-words and spaces which would reduce the performance of the classifier function. Also the Italian language is unsuitable to be run on Weka or any other preprocessing tool. Hence conversion was done to obtain the dataset as shown in figure 2. The dataset was therefore passed into a preprocessing tool that is open source in order to achieve the required quality in the dataset. After the preprocessing step the dataset would contain a continuous string of words that would the comments of the users. This process is very important as the additional data like commas and spaces would reduce the effect of the algorithm on the dataset. Thus proper care should be taken to perform the preprocessing to the correct amount.
\subsection{Classification}
Using the formula for the posterior probability the code was designed to understand the classes that the comments and users were assigned to. The classes are neutral, offensive, sexual, hatred, pun-intended. So according to the presence of these types of comments in the document, the values can be calculated by applying the algorithm. This would be used to train the system into classifying further newer comments or users into the above class categories.
Few examples of classification are as shown below:
\begin{itemize}
\item I hate this woman - Hatred
\item I had a good day - Neutral
\item I want to see you without your respect - Offensive
\end{itemize}
\subsection{Prediction}
This stage is required in order to extend the system to predict the class of a new comment as it comes dynamically on the Social Media. As the new input is received the system will calculate the probability of the comment being in any of the five classes. This prediction will automate the system and the Social Network Manager can ease his work. The comments would be classified by the algorithm and the training set and thus a successful system to filter the messages on Social Media is implemented.

\section{EXPERIMENTS}
The dataset was tested against the designed classifier to come up with the experimental results. The stage of preprocessing is done first. The results of the correctly classified and incorrectly classified instances were collected. The below table shows the same.
\begin{figure}
\centering
\includegraphics[width=3in,height=1.5in]{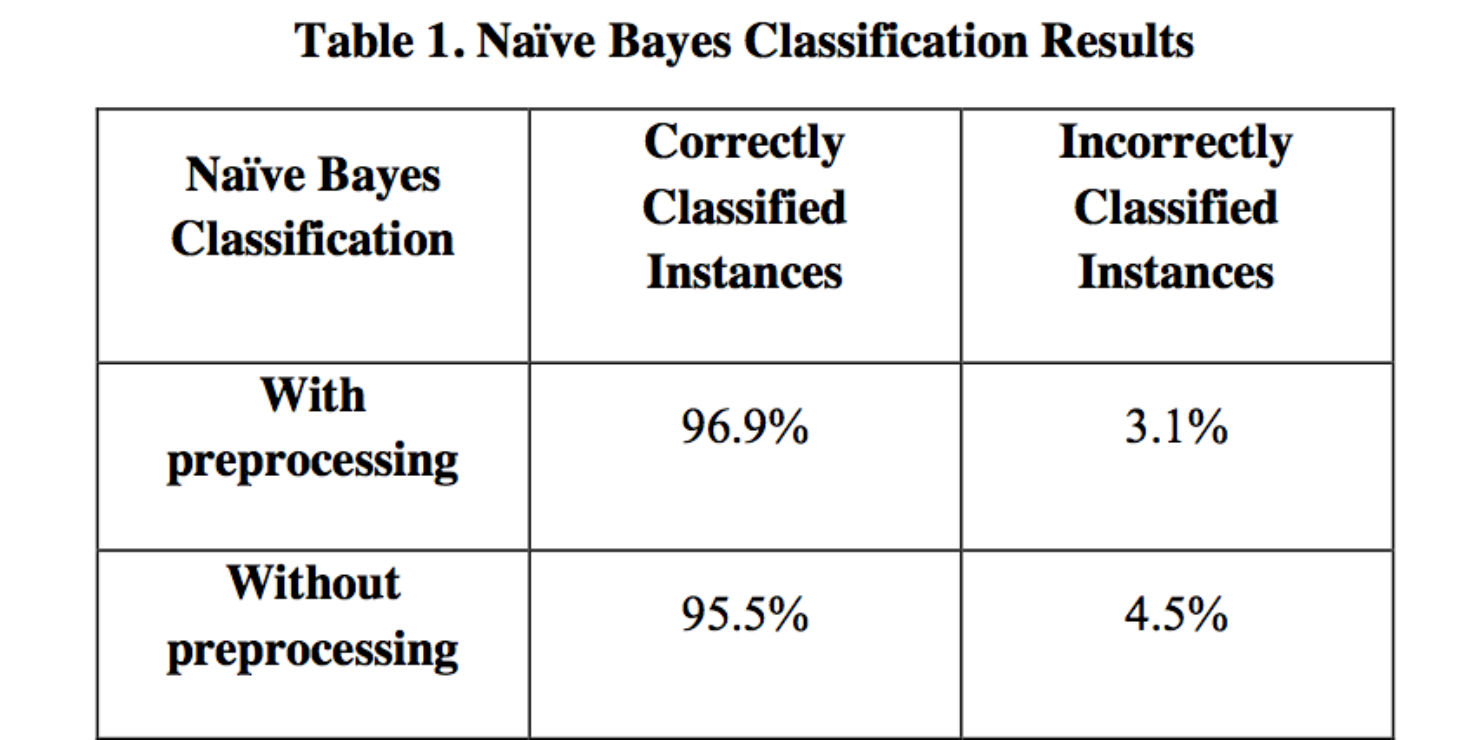}
\end{figure}
\begin{figure}
\centering
\includegraphics[width=3in,height=1.5in]{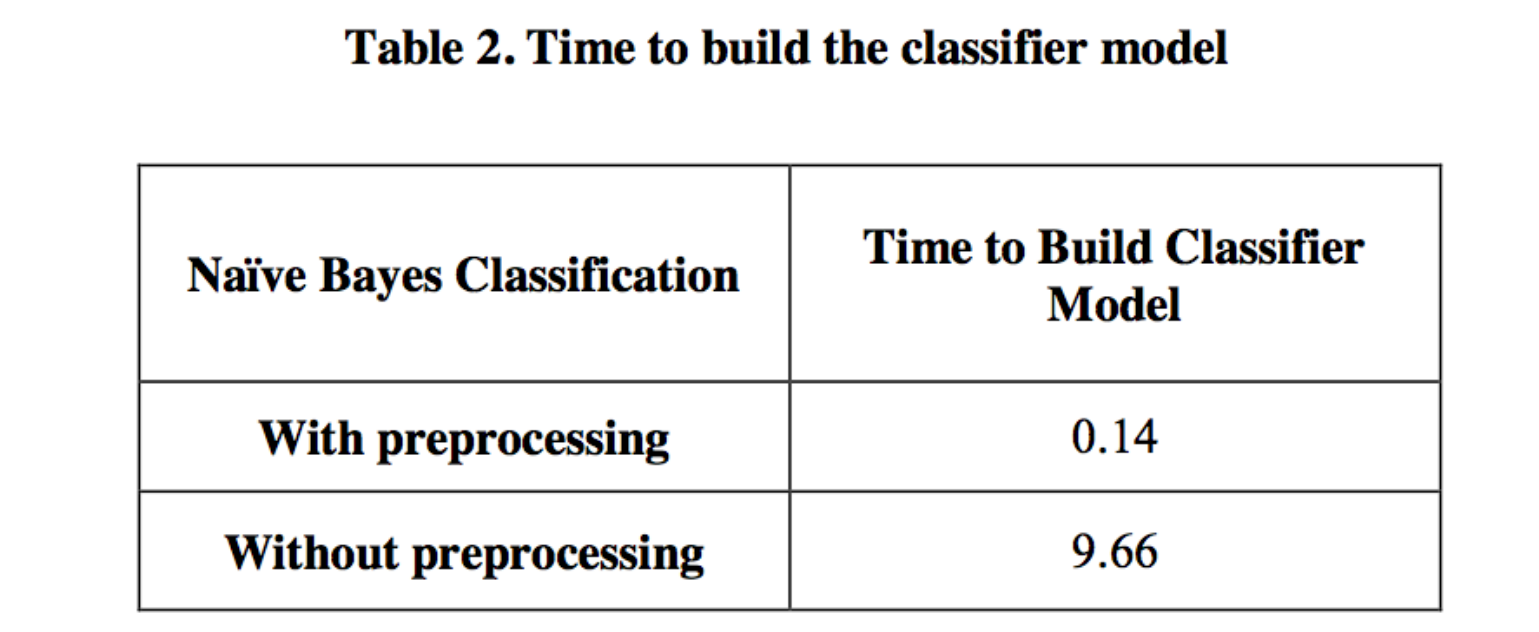}
\end{figure}
The system designed for classification is successfully running on the dataset. The time the classification had taken and to finish its task is when the model is ready for analysis. This time also varied for when the dataset was used with and without preprocessing \cite{REF02}.
\begin{figure}
\centering
\includegraphics[width=3in,height=1.5in]{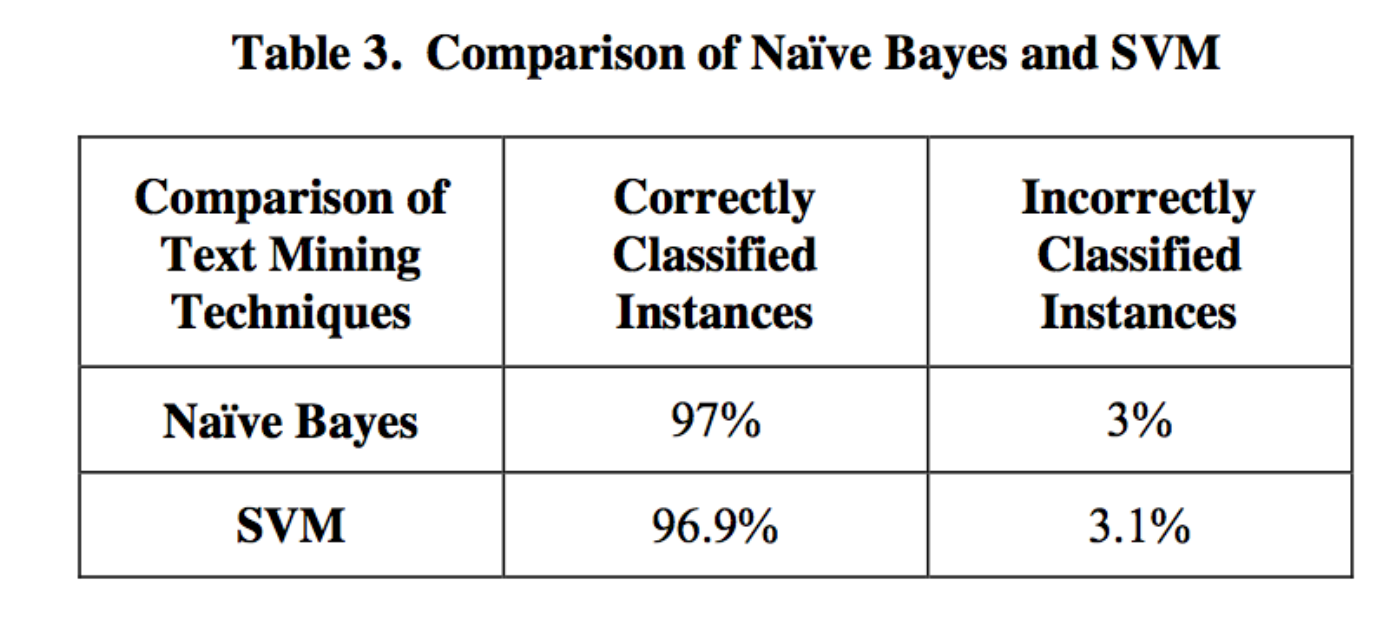}
\end{figure}
The system designe
The classifier that was built using the concept of the Naïve Bayes theorem was also compared along with another text mining technique called the Support Vector Machine \cite{REF04}. SVM is also used for text classification and clustering by the research world. Kernel support vector machines are popular tool to improve the accuracy \cite{dg1,dg2,dg3}. This comparison is needed to as to prove that the system designed and coded based of Naïve Bayes is indeed better that other option in the text mining area. The same dataset was run on both the designed Naïve Bayes Classifier and the SVM classifier to obtain the results as depicted below.
\begin{figure}
\centering
\includegraphics[width=3in,height=1.5in]{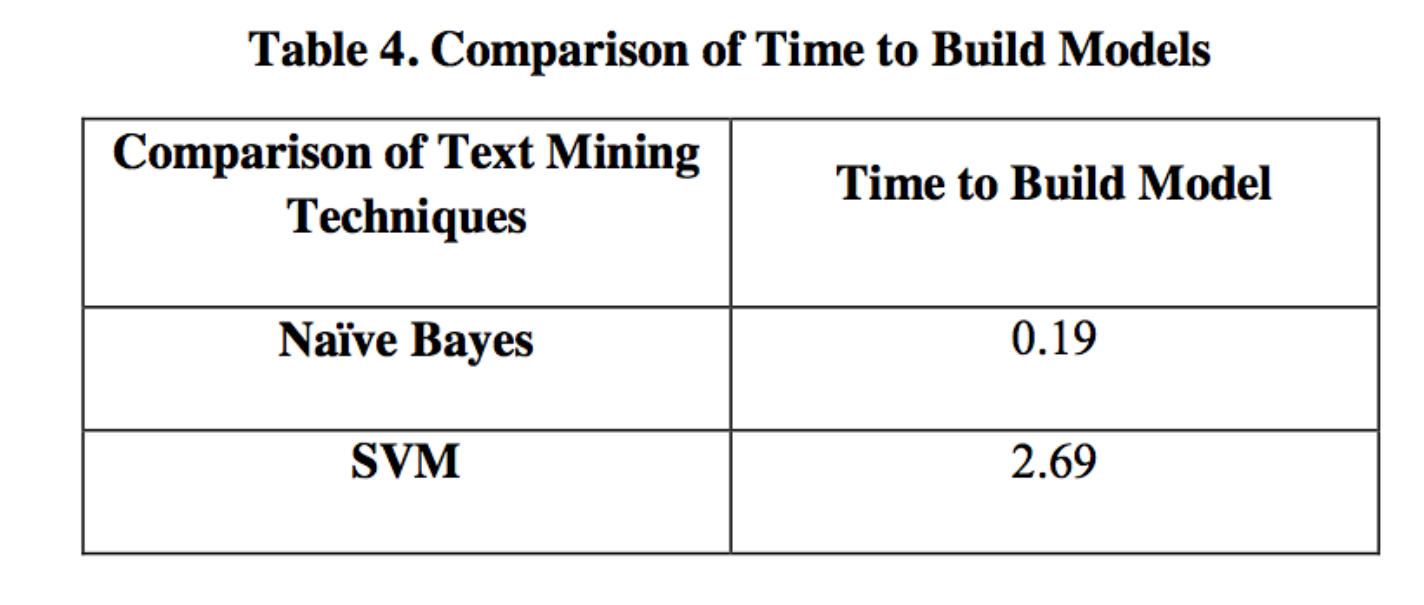}
\end{figure}
The comparison is also done to find the time taken for both the systems to run classifier models.

\section{Conclusion}
The conclusion that is drawn from the paper is that the messages that could cause harm to the image of a Social Media User can be reduced by applying this system. The messages are classified to give a better insight as to which messages or posts are to be allowed on the user walls. The Naïve Bayes theorem is used to design the classifier which can later be trained to predict the class of a new instance.
The comparison of the designed system was done with the SVM technique and was found to be better in ways of classification and performance. Thus Naïve Bayes technique is useful to classify text. Since the content is analyzed and classified it ventures into newer fields of research which would prove to be a breakthrough in the field of text mining.

{\small
\bibliographystyle{ieee}
\bibliography{egbib}
}

\end{document}